\documentclass[slac_one]{revtex4}
\usepackage{graphicx}
\usepackage{fancyhdr}
\def\pt{\ensuremath{p_\mathrm{T}}}

\def\nch{\ensuremath{n_{\mathrm{ch}}}}

\def\meanpt{\ensuremath{\langle \pt \rangle}}

\newcommand{\ptlead}{\ensuremath{\pt^\mathrm{lead}}}

\begin{document}

\title{Measurement of the properties of inelastic p-p events with the ATLAS detector} %% Paper title goes here


\author{Maaike Limper, on behalf of the ATLAS collaboration}
\affiliation{University of Iowa, 203 Van Allen Hall, Iowa City (USA)}

\begin{abstract}
New measurements are presented from proton-proton collisions at $\sqrt{s}$~=~7~TeV recorded with the ATLAS detector at the LHC. Minimum bias distributions are measured in distinct phase-space regions and compared with Monte Carlo model predictions. Activity in the underlying event is measured with respect to the highest \pt\ track in the event. Angular correlations between charged particles are studied to provide model-sensitive measurements.
\end{abstract}

\maketitle

\thispagestyle{fancy}


\vspace*{-.4cm}
\section{Introduction} 
Proton-proton interactions are composed of non-diffractive and diffractive processes that predominantly involve low momenta transfers which are not calculable within perturbative QCD and are only described by phenomenological models implemented in Monte Carlo event generators. Measurements of inclusive charged-particle distributions at previous hadron collider experiments have been used to constrain the phenomenological models of soft-hadronic interactions, providing a basis of comparison for the measurements presented in these proceedings. The LHC collision energy at $\sqrt{s}$~=~7~TeV provides a new energy regime in which properties of $pp$ collisions can be studied.

The ATLAS measurements are fully corrected for detector effects to obtain distributions at the hadron level. More detail on the event and track selection criteria and the procedure to correct for detector effects can be found in~\cite{MinBias_new}.

%The relative contribution from diffractive and non-diffractive processes depends on the choice of event and track selection. For this reason distinct regions of phase-space were defined in which the ATLAS measurements were made.
%ATLAS measured charged particle distributions in the following phase-space regions: 
%1. Events with at least 1 primary charged particle with pT>0.5 GeV and |'|<2.5
%2. Events with at least 2 primary charged particles with pT>0.1 GeV and |'|<2.5
%A considerably larger diffractive component is present when considering tracks above 100 MeV than for tracks above 500 MeV. According to both PYTHIA 6 and 8, the diffractive contribution is ~21-22% for phase-space 2, compared to ~14\% for phase-space 1.

\vspace*{-.4cm}

\vspace*{-.4cm}
\section{Charged particle distributions}
The relative contribution from diffractive and non-diffractive processes depends on the choice of event and track selection, and model predictions for the diffractive component vary greatly. In order to make model-independent measurements, no subtraction of the single-diffractive component was applied and instead ATLAS measured charged particles distributions in the following distinct phase-space regions:\\
%\begin{enumerate}
~1. Events with at least 1 primary charged particle (\nch\ $\geq$~1) with \pt\ $>$~0.5~GeV and $|\eta| <$~2.5. \\
~2. Events with at least 6 primary charged particles (\nch\ $\geq$~6) with \pt\ $>$~0.5~GeV and $|\eta| <$~2.5. \\
~3. Events with at least 2 primary charged particles (\nch\ $\geq$~2) with \pt\ $>$~0.1~GeV and $|\eta| <$~2.5.

A considerably larger diffractive component is present when considering tracks above 0.1 GeV than for tracks above 0.5 GeV. According to both PYTHIA 6 and 8, the diffractive contribution is 21-22\% for phase-space 1, compared to 14\% for phase-space 3. The event requirements for phase-space 2 reduces the diffractive contribution to a negligible amount for any model prediction.

\begin{figure}[htb!]
	\includegraphics[width=0.26\textwidth]{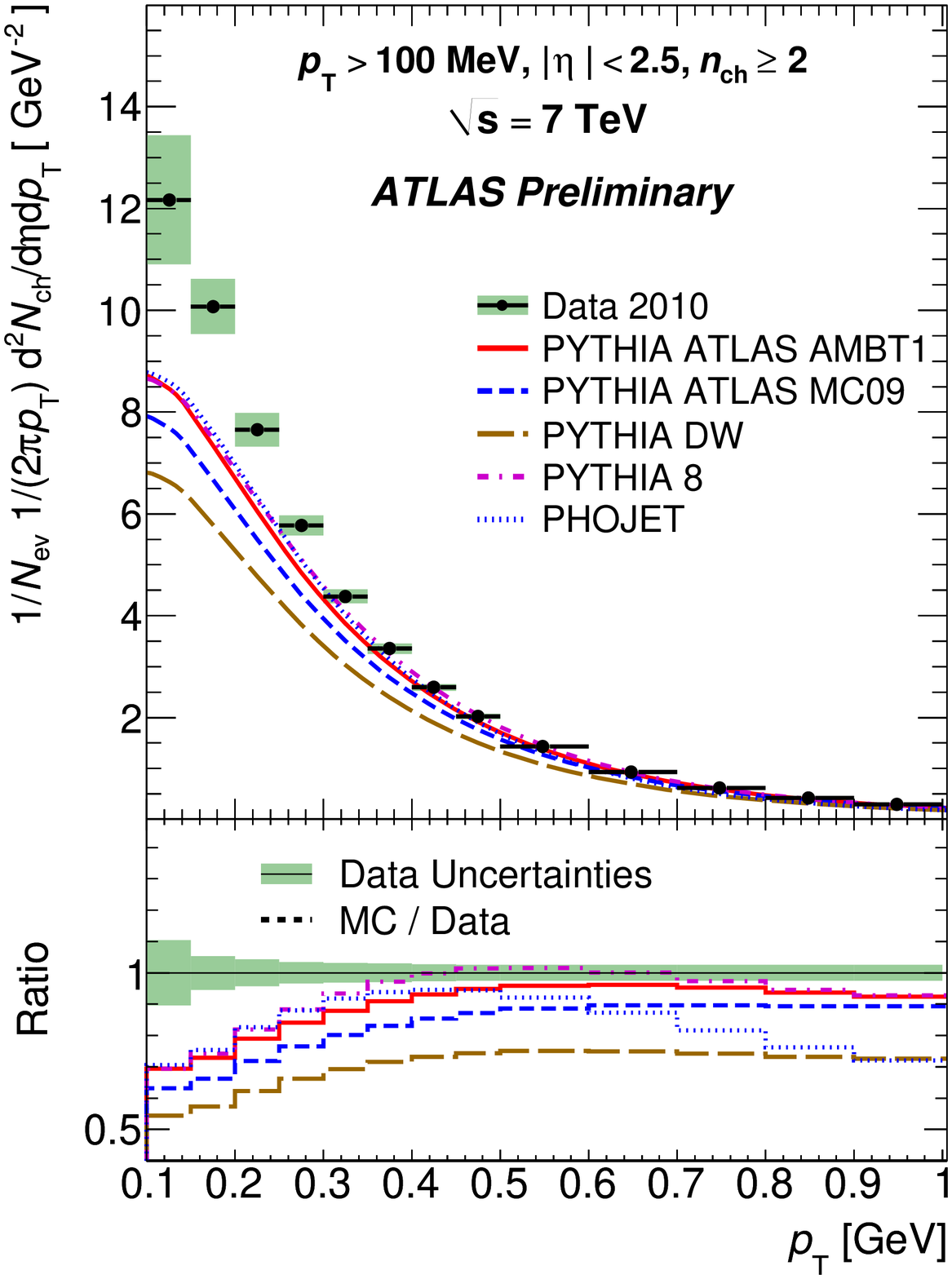}
	\includegraphics[width=0.26\textwidth]{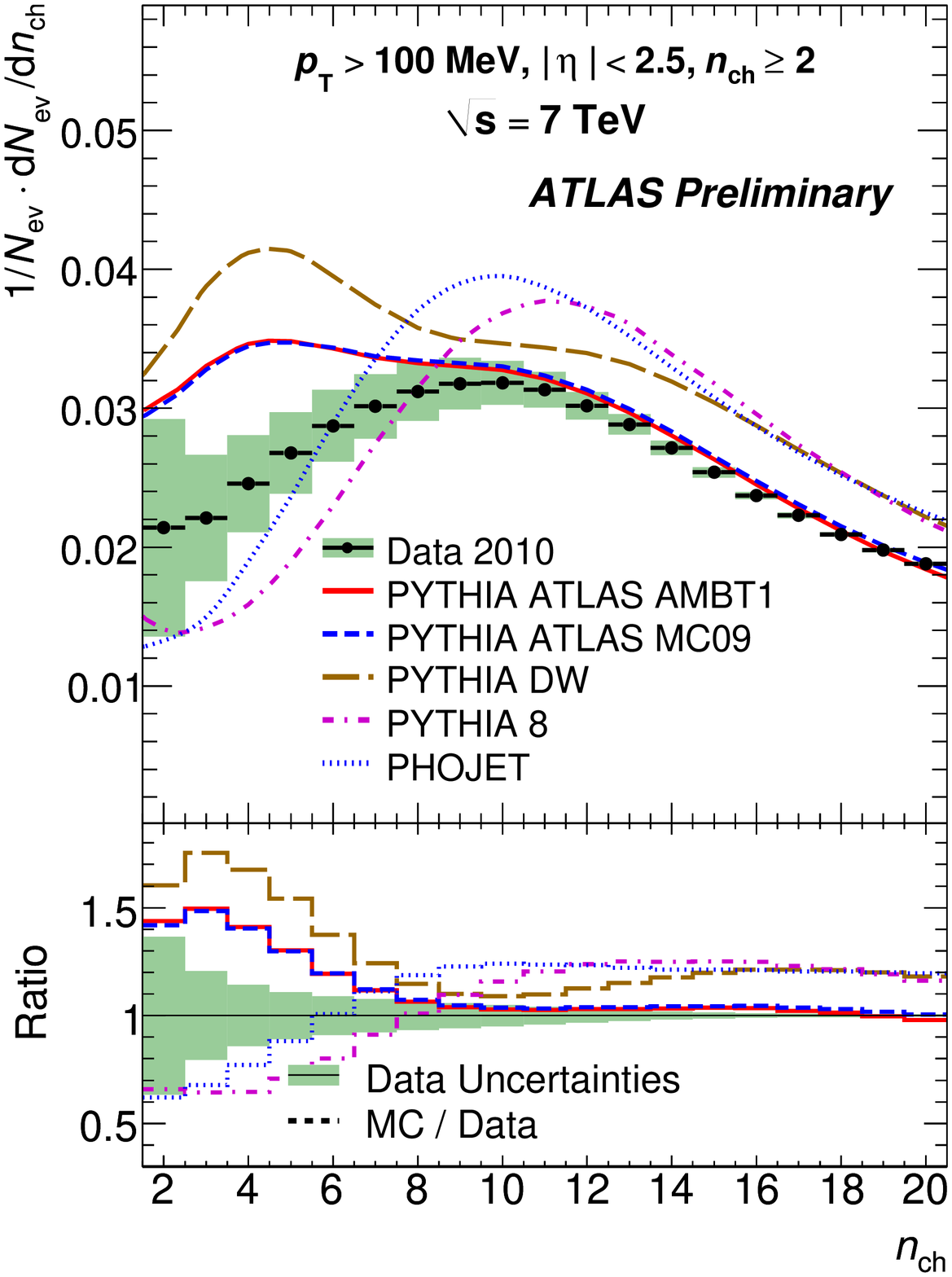}
	\includegraphics[width=0.26\textwidth]{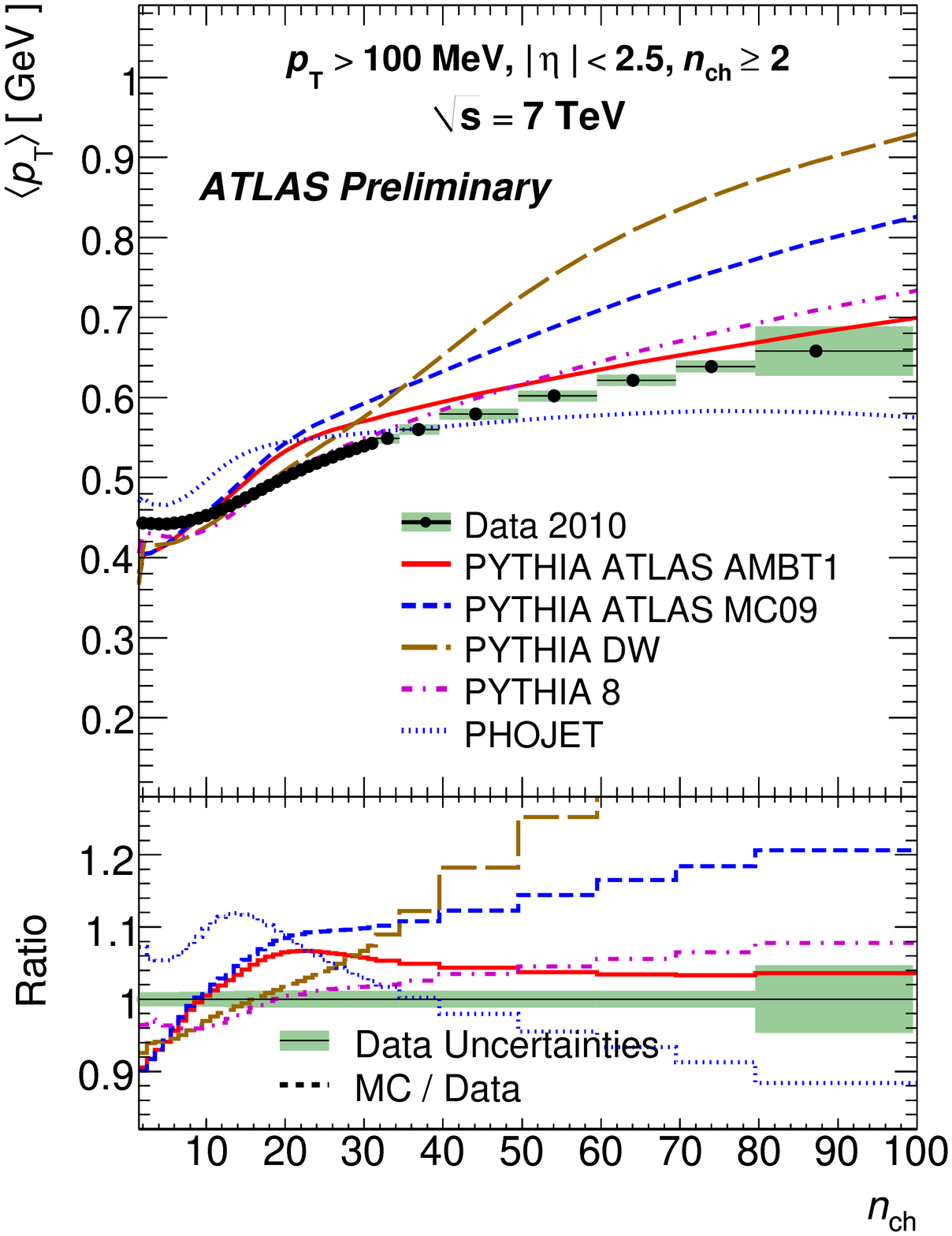}
\vspace*{-.4cm}
\caption{Charged particle distributions at $\sqrt{s}$~=~7~TeV, showing charged particle multiplicity versus \pt\ (left), charged particle multiplicity versus \nch (middle) and \meanpt\ as a function of \nch\ (right)~\cite{MinBias_new}.}
\label{MB_plots}
\end{figure}

Figure \ref{MB_plots} shows the measured charged particle distributions in phase-space 3 at $\sqrt{s}$~=~7~TeV, compared to predictions from various Monte Carlo models~\cite{MinBias_new}. The only model that was tuned using data at $\sqrt{s}$~=~7~TeV is PYTHIA ATLAS AMBT1 (ATLAS Minimum Bias Tune 1)~\cite{AMBT1}. This tune used the diffractive suppressed ATLAS measurements from phase-space 2 to improve the tuning of model parameters for non-diffractive events, leading to the relatively good agreement with the results shown in figure~\ref{MB_plots} at high \nch\ and high \pt.

%The results in \ref{MB_plots} show that the measurements agree reasonably well with ATLAS MBT1 at high \nch\ and high \pt (corresponding to the regions dominated by non-diffractive processes) while for low \pt and low \nch, the data/MC comparison is in general worse.

The measurement of charged particle distributions down to \pt =~0.1~GeV can be used to improve the model predictions for very soft particle production, similar to the way AMBT1 has helped to improve the description for the non-diffractive component.

\vspace*{-.4cm}
\section{Track based underlying event measurements}
The activity accompanying the hard scattering process, the \emph{underlying event}, was measured by looking at charged particle density and the angular distribution with respect to the highest \pt\ track (=leading track) in the event~\cite{UE_conf_note}. Activity in the underlying event is characterized by tracks measured in the transverse region, defined by azimuthal track angle with respect to the leading track of 60$^{\circ}<|\Delta \phi|<$120$^{\circ}$. 

\begin{figure}[htb!]
	\includegraphics[width=0.37\textwidth]{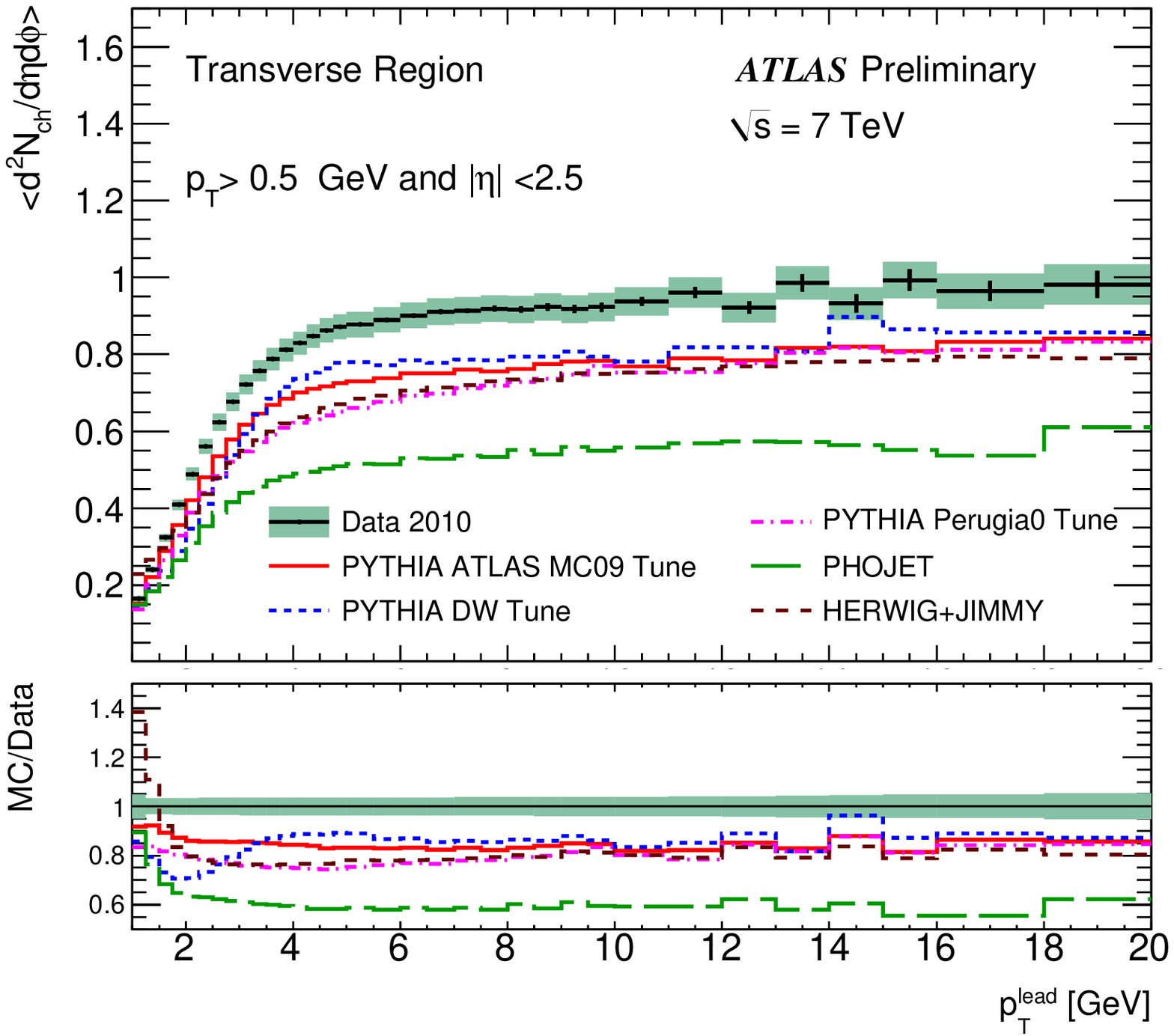}
	\includegraphics[width=0.37\textwidth]{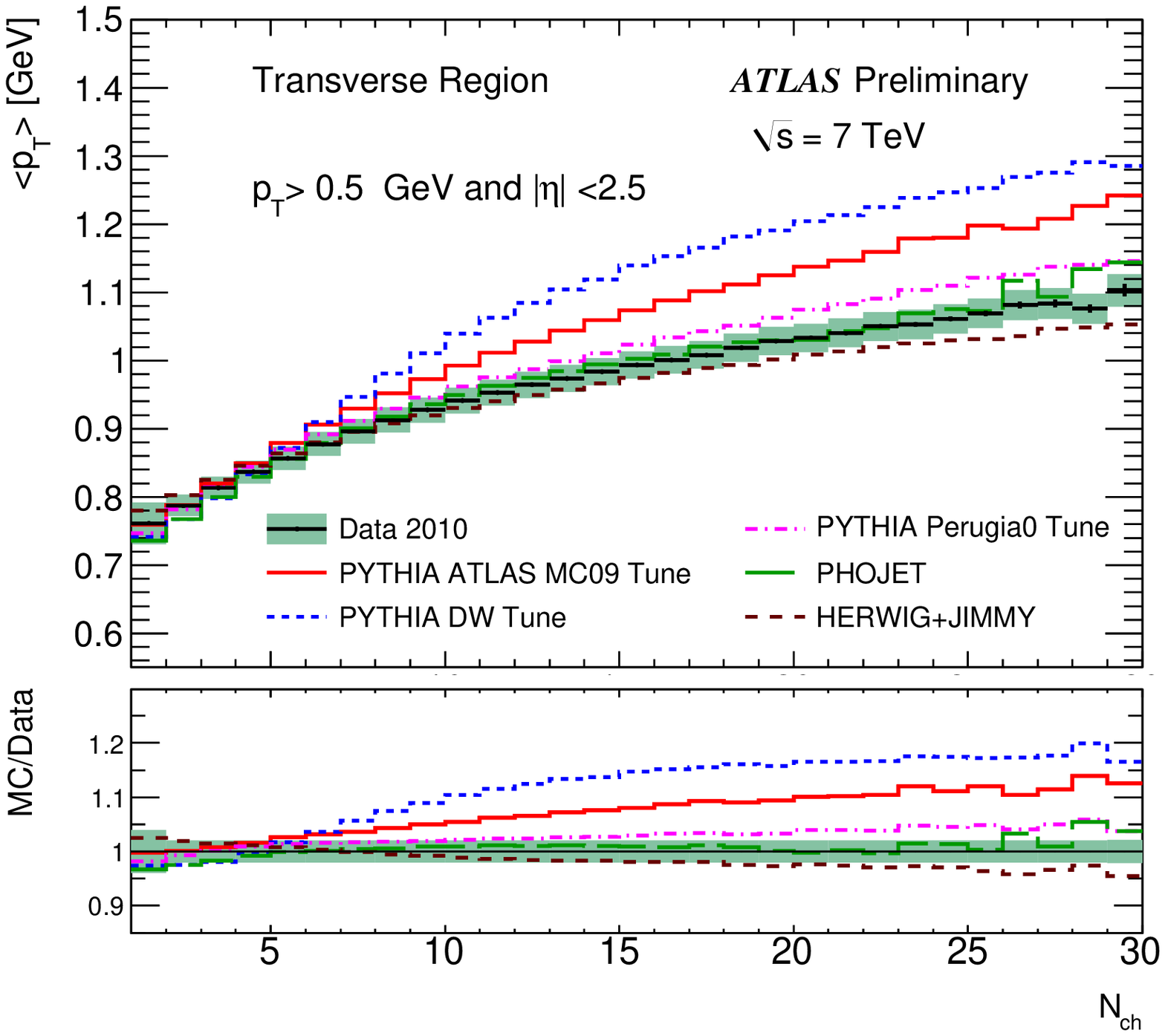}
\vspace*{-.4cm}
\caption{Distributions for the underlying event activity at $\sqrt{s}$~=~7~TeV, showing the density of charged particles as a function of \ptlead\ in the transverse region (left) and the average $\pt$ as a function of charged particle multiplicity (right)~\cite{UE_conf_note}.}
\label{UE_plots}
\end{figure}

Figure~\ref{UE_plots} shows the measured density of charged particles as a function of \ptlead\ in the transverse region and the measured average \pt\ as a function of charged particle multiplicity. The density of charged particles reaches a plateau for \ptlead\ $>$~5~GeV and shows a higher underlying event activity than predicted by any of the models, with the best prediction being given by the PYTHIA DW tune. The correlation between average \pt\ and charged particle multiplicity is a measure of the amount of hard (pertubative) and soft (non-pertubative) QCD processes in the underlying event and is best described by the PYTHIA Perugia0 tune. 

The results show that different models are better in describing different aspects of the underlying events indicating that the model predictions could be improved by using data from the underlying event measurements in ATLAS.

\vspace*{-.4cm}
\section{Angular correlations between charged particles}
The angular distribution of particles with respect to the leading particle in the transverse plane, $\Delta \phi$, provides model-sensitive observables with a relatively small systematic uncertainty. Angular correlations are characterized by the \emph{crest shape} distribution, obtained from the $\Delta \phi$-distribution and subtracting the minimum value of the distribution, and the \emph{same minus opposite} distribution, obtained by subtracting the $\Delta \phi$-distribution of particles with the opposite $\eta$-sign as the leading particle from the $\Delta \phi$-distribution of particles with the same $\eta$-sign.

\begin{figure}[htb!]
	\includegraphics[width=0.28\textwidth]{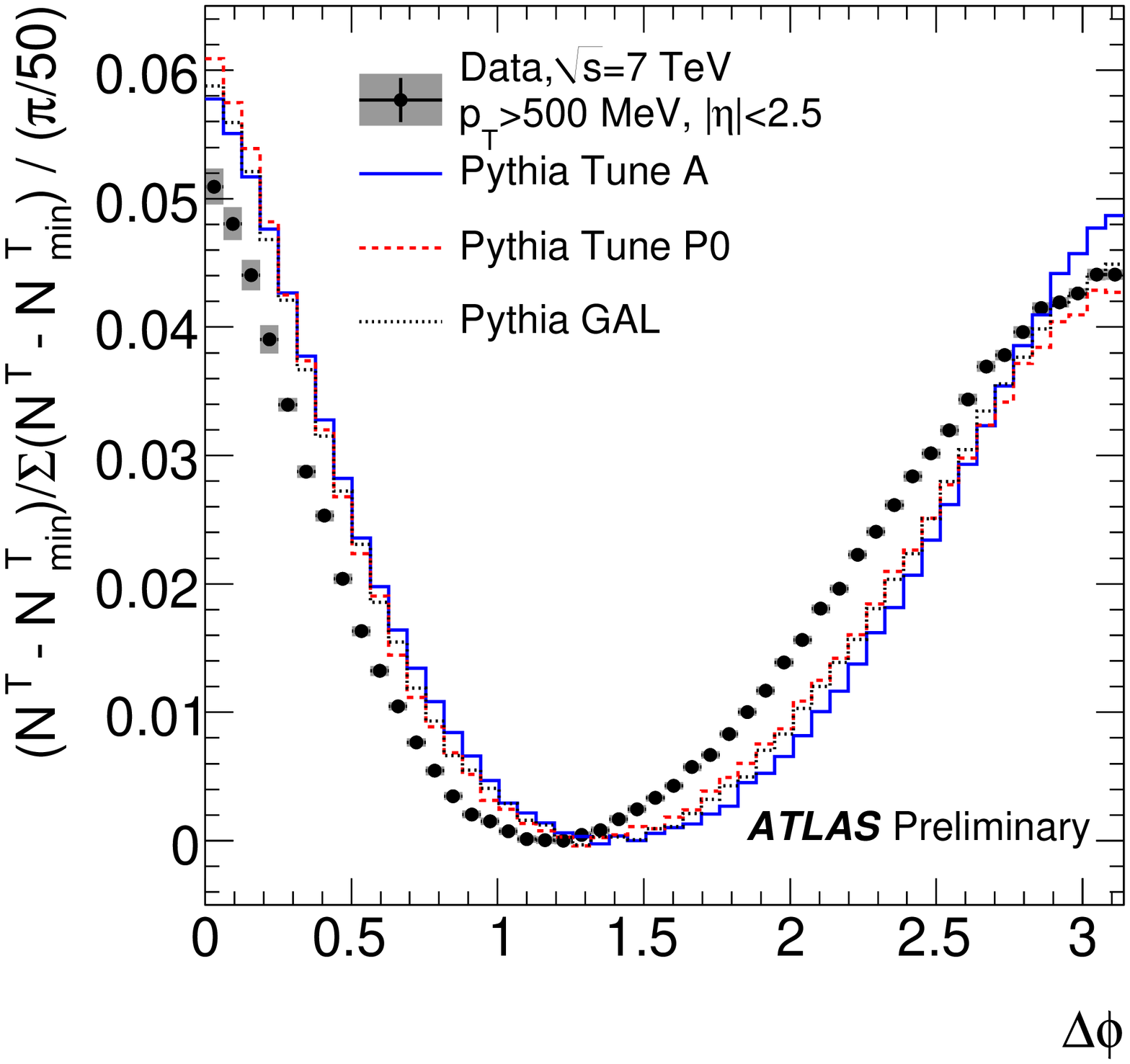}
	\includegraphics[width=0.28\textwidth]{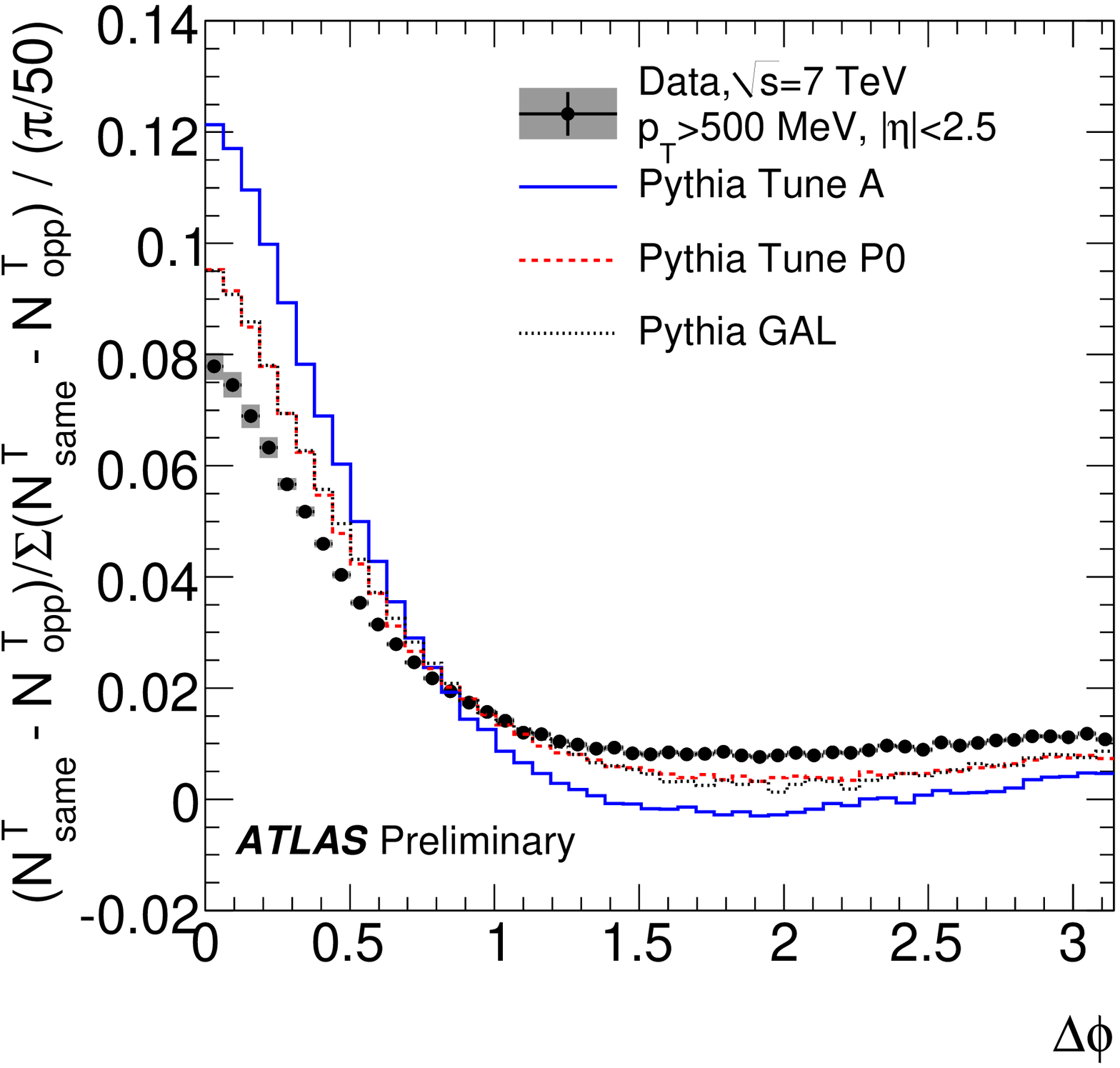}
\vspace*{-.4cm}
\caption{Observables characterizing the angular correlations measured in ATLAS events at $\sqrt{s}$~=~7~TeV, showing the ``crest shape'' distribution (left) and ``same minus opposite'' distribution (right)~\cite{angular_conf_note}.}
\label{angular_plots}
\end{figure}

Figure~\ref{angular_plots} shows the results for the measured observables in ATLAS for particles with \pt\ $>$~500~MeV and $\eta <$~2.5~\cite{angular_conf_note}.  To illustrate the dependence of the model parameters and techniques, the measured observables are compared with various MC models. PYTHIA Tune A is an older tune which uses virtuality-ordered showers, Perugia0 (P0) is a more recent tune that uses \pt-ordered showers and the PYTHIA GAL tune is similar to P0 in most aspects, except its use of an alternative color reconnection model. Comparison of the curves show that the observables are sensitive to these different models descriptions and that none of the models is capable of correctly describing the curves measured in data, indicating that the measurements provide useful information to improve the model description.
% the Generalized Area Law (GAL) model .

\vspace*{-.4cm}
\section{Conclusion}
The ATLAS detector at the LHC measured various properties of inelastic proton-proton interactions. Charged particle distributions in minimum bias events were measured in distinct phase-space regions. Both the average charged particle multiplicity in minimum bias events and the activity in the underlying event were measured to be significantly above the predictions of any of the Monte Carlo models.
%Activity in the underlying event was measured with respect to the highest \pt\ track in the event. 

The measurements of charged particle distributions, underlying event activity and angular correlations between charged particles provide useful information to improve the phenomenological models that describe inelastic pp interactions in the new energy regime set by LHC.


\begin{thebibliography}{9}   % Use for  1-9  references

%\bibitem{ATLAS}
%{The ATLAS} Collaboration, {\em {The ATLAS Experiment at the CERN Large Hadron Collider}\/},
%{JINST {\bf 3} (2008) S08003}.
%%CITATION = JINST,3,S08003;%%.

\bibitem{MinBias_new}
{The ATLAS} Collaboration, {\em {Charged particle multiplicities in pp interactions for track \pt\ $>$~100MeV at $\sqrt{s}$ = 0.9 and 7 TeV measured with the ATLAS detector at the LHC}\/} (2010).
\newblock ATLAS conference note: ATL-CONF-2010-046.

\bibitem{AMBT1}
{The ATLAS} Collaboration, {\em {Charged particle multiplicities in pp interactions at $\sqrt{s}$ = 900 GeV and 7 TeV in a diffractive limited phase-space measured with the ATLAS detector at the LHC and new PYTHIA6 tune}\/} (2010).
\newblock ATLAS conference note: ATL-CONF-2010-031.

\bibitem{UE_conf_note}
{The ATLAS} Collaboration, {\em {Track-based underlying event measurements in pp collisions at $\sqrt{s}$ = 900 GeV and 7 TeV with the ATLAS detector at the LHC}\/} (2010).
\newblock ATLAS conference note: ATL-CONF-2010-081.

\bibitem{angular_conf_note}
{The ATLAS} Collaboration, {\em {Angular correlations between charged particles from proton-proton collisions at $\sqrt{s}$ = 900 GeV and 7 TeV measured with the ATLAS detector}\/} (2010).
\newblock ATLAS conference note: ATL-CONF-2010-082.

\end{thebibliography}
\end{document}